\documentclass[amsmath,amssymb,reprint,aps,pra,floatfix,superscriptaddress]{revtex4-2}
\usepackage{bm}
\usepackage{amsmath}
\usepackage{graphicx,float}
\usepackage{braket}
\usepackage{bbm}
\usepackage[hidelinks]{hyperref}

\begin{document}

\title{Experimental demonstration of Gaussian boson sampling with displacement}

\author{G.S. Thekkadath}
\thanks{These two authors contributed equally.}
\affiliation{Department of Physics, Imperial College London, Prince Consort Rd, London SW7 2AZ, UK}
\affiliation{National Research Council of Canada, 100 Sussex Drive, Ottawa, K1A 0R6, Canada}

\author{S. Sempere-Llagostera}
\thanks{These two authors contributed equally.}
\affiliation{Department of Physics, Imperial College London, Prince Consort Rd, London SW7 2AZ, UK}

\author{B.A. Bell}
\affiliation{Department of Physics, Imperial College London, Prince Consort Rd, London SW7 2AZ, UK}

\author{R.B. Patel}
\affiliation{Department of Physics, Imperial College London, Prince Consort Rd, London SW7 2AZ, UK}
\affiliation{Clarendon Laboratory, University of Oxford, Parks Road, Oxford, OX1 3PU, UK}

\author{M.S. Kim}
\affiliation{Department of Physics, Imperial College London, Prince Consort Rd, London SW7 2AZ, UK}

\author{I.A. Walmsley}
\affiliation{Department of Physics, Imperial College London, Prince Consort Rd, London SW7 2AZ, UK}

\begin{abstract}
Gaussian boson sampling (GBS) is quantum sampling task in which one has to draw samples from the photon-number distribution of a large-dimensional nonclassical squeezed state of light.
In an effort to make this task intractable for a classical computer, experiments building GBS machines have mainly focused on increasing the dimensionality and squeezing strength of the nonclassical light.
However, no experiment has yet demonstrated the ability to displace the squeezed state in phase-space, which is generally required for practical applications of GBS.
In this work, we build a GBS machine which achieves the displacement by injecting a laser beam alongside a two-mode squeezed vacuum state into a 15-mode interferometer.
We focus on two new capabilities.
Firstly, we use the displacement to reconstruct the multimode Gaussian state at the output of the interferometer.
Our reconstruction technique is \textit{in situ} and requires only three measurements settings regardless of the state dimension.
Secondly, we study how the addition of classical laser light in our GBS machine affects the complexity of sampling its output photon statistics.
We introduce and validate approximate semi-classical models which reduce the computational cost when a significant fraction of the detected light is classical.
\end{abstract}

\maketitle

\section{Introduction}
Several recent experiments employing large-scale quantum systems entered a complexity regime where they cannot currently be simulated on a classical computer~\cite{arute2019quantum,zhong2020quantum,zhong2021phase,wu2021strong}.
These experiments reached a milestone on the path to using quantum systems for solving computational tasks of practical importance that are intractable for classical computers~\cite{preskill2018quantum}. 
One of the approaches used to reach this milestone is called Gaussian boson sampling (GBS) and consists in injecting a large number of nonclassical squeezed states of light into a multiport interferometer~\cite{hamilton2017gaussian}.
Light at the output of the interferometer is generally in a complex entangled state owing to quantum interference.
This output state is then measured using an array of single-photon detectors.
The complexity of calculating the output light photon statistics scales with the number of interferometer modes and the number of detected photons~\cite{aaronson2011computational}.
Over the years, these numbers have been increasing in part thanks to improvements in the quality and brightness of the squeezed light sources~\cite{harder2016single,zhong2018twelve,vaidya2020broadband}, the efficiency and energy resolution of the detectors~\cite{lita2008counting,marsili2013detecting}, as well as the development of scalable chip-based experiments~\cite{paesani2019generation,bell2019testing,arrazola2021quantum}.
The largest GBS experiment performed thus far employed 25 squeezed light sources and measured over 100 photons at the output of a 144-mode interferometer~\cite{zhong2021phase}.


Although a GBS machine is not a universal quantum computer, drawing samples from the output photon  distribution has several potential applications, including calculating the vibronic spectra of molecules~\cite{huh2015boson} and characterizing features of graphs~\cite{bradler2018gaussian,arrazola2018using,schuld2020measuring,banchi2020molecular}.
These applications require the ability to control the squeezing and displacement of the input light as well as program the interferometer transformation.
Experiments have already demonstrated the ability to implement arbitrary transformations by using reconfigurable multiport interferometers~\cite{carolan2015universal,sparrow2018simulating,taballione20198,arrazola2021quantum}.
In fact, updating such transformations in a feedback loop based on measurement outcomes has been used for machine learning~\cite{saggio2021experimental} and variational quantum algorithms~\cite{peruzzo2014variational}.
However, no experiment has yet demonstrated the ability to displace the squeezed light in a GBS machine.
Displacements can improve the graph classification accuracy of GBS~\cite{schuld2020measuring} and are needed for calculating the vibronic spectra of real molecules~\cite{huh2015boson,huh2017vibronic,wang2020efficient}.
Moreover, a GBS machine equipped with displacements provides a powerful quantum state engineering tool that can conditionally prepare arbitrary single-mode states~\cite{dakna1999generation,fiurasek2005conditional,su2019conversion,tzitrin2020progress}.%

The displacement can be achieved by interfering the nonclassical squeezed light with laser light, i.e. a coherent state.
Ref.~\cite{afek2010high} experimentally demonstrated that the interference between these two states can be used for quantum-enhanced interferometry.
Other experiments observed nonclassical features of displaced quantum states such as oscillations in the photon-number distribution~\cite{schleich1987oscillations,lvovsky2002synthesis} and micro-macro entanglement~\cite{lvovsky2013observation}.
In the context of GBS, the addition of laser light provides an easy way to increase the photon detection rate and thus reduce statistical errors when sampling the output photon-number distribution.
However, it also raises questions regarding the complexity of the GBS problem.
If the laser is much brighter than the squeezed light, then one might be able to find a classical approximation for the output photon distribution that can be efficiently sampled using classical algorithms~\cite{Seshadreesan2015boson,rahimi2016sufficient,qi2020regimes}.
Thus, the GBS machine would no longer provide a quantum computational advantage.
It is not well understood where this transition in complexity occurs in practice.


In this work, we build a GBS machine that samples from a 15-dimensional displaced Gaussian state.
Our experiment employs a single source of squeezed vacuum and thus it can be readily  simulated on a classical computer.
Rather than aiming to achieve a quantum computational advantage, we explore two new capabilities enabled by the displacement. 
Firstly, we show that it can be used to determine the multimode Gaussian state at the output of the circuit, i.e. perform a high-dimensional quantum state reconstruction.
Secondly, we study the complexity of simulating GBS machines in the presence of displacements.
We introduce approximate semiclassical models that reduce the computational cost of simulations.
Similar to a quantum-to-classical transition, we find that the validity of these models generally improves as we increase the displacement strength. 

\section{Theory}
\begin{figure}
    \centering
    \includegraphics[width=0.5\columnwidth]{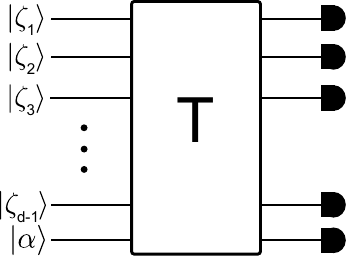}
	\caption{
	Gaussian boson sampling problem considered. Squeezed vacua $\ket{\zeta_i}$ and a single coherent state $\ket{\alpha}$ are injected into a $d$-mode lossy interferometer described by a transfer matrix $\bm{T}$.
	The coherent state displaces the phase-space distribution of the output state.
	}
	\label{fig:gbs_schematic}
\end{figure}
Consider injecting squeezed vacua states $\ket{\zeta_i}$ and a single-mode coherent state $\ket{\alpha}$ ($\alpha = |\alpha|e^{i\phi})$ into separate ports of a $d$-mode lossy interferometer, as shown in Fig.~\ref{fig:gbs_schematic}.
Light at the output of the interferometer is in a Gaussian state that can be fully described by a $2d \times 2d$ covariance matrix $\bm{\Sigma}$ and a displacement vector $\bm{\delta}$ of length $2d$:
\begin{subequations}
\begin{align}
    \Sigma_{j,k} &= \frac{1}{2}\left ( \braket{\nu_j \nu^\dagger_k} + \braket{\nu^\dagger_k \nu_j} \right) - \delta_j\delta_k^* , \label{eqn:covariance} \\
    \delta_j &= \braket{\nu_j}.
\label{eqn:displacement_vec}
\end{align}
\end{subequations}
Here $\bm{\nu}=(\hat{a}_1, ..., \hat{a}_d, \hat{a}^\dagger_1,...,\hat{a}^\dagger_d)$ is a vector of boson annihilation and creation operators.
The former quantity describes the squeezing and thermal occupation of each mode (and their correlations) after propagating the input squeezed light through the interferometer.
The latter quantity provides the displacement amplitude of each mode and is determined by the evolution of the coherent state.
We consider all losses to be part of the interferometer circuit, i.e. $\bm{\Sigma}$ and $\bm{\delta}$ define the state just before being measured by ideal detectors.
We use the convention that uppercase (lowercase) bold symbols are matrices (vectors).

GBS experiments performed thus far have not employed displacements, i.e. $\bm{\delta} =0$.
In this case, the probability to obtain the detection pattern $\bm{n}=(n_1,...,n_d)$ is given by~\cite{hamilton2017gaussian}
\begin{equation}
    \mathrm{pr}(\bm{n}) = \frac{p_{\mathrm{vac}}}{\prod_i n_i!} \times \mathrm{haf}(\bm{A}_{\bm{n}}) 
    \label{eqn:gbs_haf}
\end{equation}
where $p_{\mathrm{vac}}$ is the probability to measure vacuum in all output modes.
We introduced the $\bm{A}$ matrix:
\begin{equation}
\label{eqn:A_matrix}
    \bm{A} = \begin{pmatrix}
0 & \bm{I}_{d}\\
\bm{I}_{d} & 0 
\end{pmatrix}
(\bm{I}_{2d} - \bm{\Sigma}_Q^{-1})
\end{equation}
with $\bm{I}_{d}$ being the identity matrix of dimension $d$ and $\bm{\Sigma}_Q = \bm{\Sigma} + \bm{I}_{2d}/2$ is the covariance matrix of the state's $Q$-function.
The submatrix $\bm{A}_{\bm{n}}$ is determined by repeating $n_i$ times the $i$th and $(i+d)$th row and column of $\bm{A}$ and thus its size ($2N \times 2N$) grows with the total number of photons detected, $N = \sum_i n_i$.
The quantity $\mathrm{haf}(\bm{A}_{\bm{n}})$ is called the hafnian.  
It sums over the product of pairs of elements in $\bm{A}_{\bm{n}}$ chosen from the set of perfect matching permutations of $2N$, which involves summing $(2N-1)!!$ terms.
This exponential scaling is at the heart of the complexity of GBS.


In contrast, when $\bm{\delta} \neq 0$, the probability to obtain a particular detection pattern is given by the loop hafnian~\cite{bjorklund2019faster}:
\begin{equation}
\mathrm{pr}(\bm{n}) = \frac{p_{\mathrm{vac}}}{\prod_i n_i!} \times \mathrm{lhaf}( \tilde{\bm{A}}_{\bm{n}} )
\label{eqn:gbs_loop_haf}
\end{equation}
which contains additional terms compared to Eq.~\eqref{eqn:gbs_haf} as it now includes the different possible ways that the photons could have originated from both the squeezers and the coherent state.
The submatrix $\tilde{\bm{A}}_{\bm{n}}$ is determined the same way as $\bm{A}_{\bm{n}}$ but its diagonal entries are replaced by $\tilde{\bm{\gamma}}$ whose elements are given by repeating $n_i$ times the $i$th and $(i+d)$th elements of 
\begin{equation}
\label{eqn:gamma}
\bm{\gamma} = \bm{\delta}^\dagger \Sigma_Q^{-1}.
\end{equation}
Expanding the loop hafnian, we find~\cite{kruse2019detailed}:
\begin{equation}
\begin{split}
    \mathrm{pr}(\bm{n}) &= \frac{p_{\mathrm{vac}}}{\prod_i n_i!}  \bigg( \mathrm{haf}(\bm{A}_{\bm{n}}) \\ &\quad+ \sum_{j_1=1}^{2N-1}\sum_{j_2=j_1+1}^{2N} \tilde{\gamma}_{j_1}\tilde{\gamma}_{j_2}\mathrm{haf}(\bm{A}_{\bm{n}-\left\{j_1,j_2\right\}}) \\
    &\quad +...+ \prod_{j=1}^{2N}\tilde{\gamma}_{j} \bigg).
\end{split}
    \label{eqn:gbs_loop_haf_expanded}
\end{equation}
The submatrix $\bm{A}_{\bm{n}-\left\{j_1,j_2\right\}}$ is obtained by removing rows and columns numbered $j_1$ and $j_2$ from $\bm{A}_{\bm{n}}$. 
The first term in Eq.~\eqref{eqn:gbs_loop_haf_expanded} is the same as Eq.~\eqref{eqn:gbs_haf} and accounts for the probability that all $N$ photons originated from the squeezers.
In contrast, the last term contains no hafnians and accounts for the probability that all photons originated from the coherent state.
The remaining terms account for the different possible ways that the $N$ photons could have originated from both sources and the interference between these possibilities.

\subsection{Reconstructing the multimode Gaussian state}
\label{sec:reconstructing_state}
In continuous-variable quantum state tomography~\cite{lvovsky2009continous}, one can reconstruct the quantum state of an unknown signal by interfering it with a coherent state and measuring the output photon statistics.
To extend this idea to multimode signals, one should interfere the coherent state with every signal mode and measure the joint photon statistics across all modes~\cite{beck2000quantum,iaconis2000temporal,beck2001joint,fabre2020modes}.
This is precisely what the GBS circuit in Fig.~\ref{fig:gbs_schematic} achieves, which raises the question of whether it is possible to reconstruct the state at the output of the circuit directly from the measured photon statistics in this configuration.
This reconstruction would provide a way to verify that the GBS machine has been properly programmed for a desired calculation.

We find that the ability to control the coherent state's phase $\phi$ can be used to determine the matrix $\bm{A}$ and the vector $\bm{\gamma}$.
The former quantity determines the covariance matrix [Eq.~\eqref{eqn:A_matrix}] and the latter determines the displacement vector [Eq.~\eqref{eqn:gamma}], thereby fully characterizing the multimode Gaussian state just before detection.
The $\bm{A}$ matrix can be written in terms of four $d \times d$ blocks, $\bm{B}$ and $\bm{C}$:
\begin{equation}
 \bm{A} = \begin{pmatrix}
\bm{B} & \bm{C}\\
\bm{C}^T & \bm{B}^* 
\end{pmatrix}.
\end{equation}
Here $\bm{B}$ ($\bm{C}$) is a symmetric (Hermitian) matrix describing the squeezed (thermal) part of the state, e.g. $\bm{C}=0$ for a pure state~\cite{hamilton2017gaussian}.
It suffices to measure single-photon outcomes, $p_j \equiv \mathrm{pr}(0_1,...,1_j,...,0_d)/p_{\mathrm{vac}}$, as well as two-photon outcomes, $p_{j,k} \equiv \mathrm{pr}(0_1,...,1_j,...,1_k,...,0_d)/p_{\mathrm{vac}}$, to determine $\bm{\gamma}$, $\bm{B}$, and $\bm{C}$, as follows.


The single-photon probabilities measured with and without the coherent state blocked are (respectively) obtained by expanding Eq.~\eqref{eqn:gbs_loop_haf}:
\begin{subequations}
\begin{align}
    p_j &= C_{j,j}, \label{eqn:singles_j_blocked} \\
    p'_j &= C_{j,j} + |\gamma_j|^2. \label{eqn:singles_j_unblocked}
\end{align}
\end{subequations}
Eq.~\eqref{eqn:singles_j_blocked} directly determines $C_{j,j}$.
These can then be used in Eq.~\eqref{eqn:singles_j_unblocked} to determine $\gamma_j$, i.e. $\gamma_j =  \sqrt{p'_j - C_{j,j}}$.
We assume that $\gamma_j$ is a real number and thus neglect the phase difference between output modes in the displacement vector.
These phases do not affect the output photon statistics since photon counters are phase insensitive.

Next, the two-photon probabilities measured with and without the coherent state blocked are (respectively) given by
\begin{subequations}
\begin{align}
    p_{j,k} &= p_jp_k + |B_{j,k}|^2 + |C_{j,k}|^2, \label{eqn:twos_j_blocked} \\
\begin{split} 
    p'_{j,k} &= p'_jp'_k + |B_{j,k}|^2 + |C_{j,k}|^2 \\ 
    &\quad + 2\gamma_j\gamma_k \left(\mathrm{Re}[C_{j,k}] + \mathrm{Re}[B^*_{j,k}e^{i2\phi}] \right).
\end{split}    
\label{eqn:twos_j_unblocked}
\end{align}
\end{subequations}
By scanning the phase $\phi$ of the coherent state, one can determine $|B_{j,k}|$ from the amplitude and $\mathrm{arg}(B_{j,k})$ from the offset of the observed fringe in $p'_{j,k}$ (i.e. the last term in Eq.~\eqref{eqn:twos_j_unblocked}).
Once $B_{j,k}$ is determined, one can solve for $|C_{j,k}|$ using Eq.~\eqref{eqn:twos_j_blocked} and subsequently $\mathrm{Re}(C_{j,k})$ using Eq.~\eqref{eqn:twos_j_unblocked}, thus only leaving an ambiguity in the sign of the imaginary part of $C_{j,k}$.
The threefold statistics can be measured to determine this sign, e.g. by employing an algorithm like maximum likelihood to minimize the distance between the measured threefolds and those calculated via Eq.~\eqref{eqn:gbs_loop_haf}.
Alternatively, we show in Appendix~\ref{sec:appendix_characterization} that one can inject the coherent state into a different input mode of the interferometer in order to determine the imaginary part of $C_{j,k}$.
Thus, a total of three measurement settings and the ability to scan the phase $\phi$ is required to fully determine the multimode Gaussian state \textit{in situ}.
These measurement settings are: (i) $\ket{\alpha}$ blocked, (ii) $\ket{\alpha}$ injected into a first input mode, (iii) $\ket{\alpha}$ injected into a second input mode.
Since $p_{j,k}=p_{k,j}$, the reconstructed $\bm{B}$ ($\bm{C}$) matrix is constrained to be symmetric (Hermitian), as required for a physical state.
Uncertainties in the measured probabilities (e.g. from counting statistics and fitting) can be propagated in Eqs.~(8) and (9) to determine the uncertainty on each matrix element.

The efficiency of our reconstruction technique is partly due to the strong constraint imposed by assuming that the output state is Gaussian.
While this is a natural assumption to make in the context of GBS, in practice, imperfections such as phase drifts in the displacement field can render the experimentally generated state non-Gaussian~\cite{allevi2013characterization,bourassa2021fast}.
Provided that these imperfections are minor, the reconstructed Gaussian state provides a good approximation of the experimentally generated state in that it accurately reproduces its photon statistics, as we demonstrate further below.

\subsection{The k-order classical approximation}
\label{sec:k-order}
The complexity of calculating $\mathrm{pr}(\bm{n})$ is determined by the largest hafnian in Eq.~\eqref{eqn:gbs_loop_haf_expanded}, whose size is determined by the number of detected photons, $N$~\cite{kruse2019detailed, bjorklund2019faster,queseda2020exact,quesada2022quadratic}.
Injecting bright coherent light into a GBS circuit is an easy way to increase the likelihood of detecting a large number of photons and thus effectively increase the $N$ achievable in an experiment.
At first glance, this does increase the difficulty of simulating the experiment because one cannot rule out the possibility that the $N$ photons originated from the squeezers.
Of course, the likelihood of this occurring depends on the relative amount of photons originating from the squeezed and coherent light, which is reflected in the weights of the different terms in Eq.~\eqref{eqn:gbs_loop_haf_expanded}.
This leads us to considering an approximate model which ignores terms that are small when the coherent light is bright relative to the squeezers.
The ``$k$-order approximation" only keeps terms in Eq.~\eqref{eqn:gbs_loop_haf_expanded} for which $\tilde{\gamma}_{j}$ appears at least $2N-2k$ times.
Roughly speaking, this assumes that at most $k$ photons originated from the squeezers.
For example, if $k=0$, we assume that all the photons came from the coherent state and only calculate $\prod_j^{2N}\tilde{\gamma}_{j}$ which contains no hafnians, whereas for $k=N$, we calculate all the terms in Eq.~\eqref{eqn:gbs_loop_haf_expanded}.
Intermediate $k$ values reduce calculation times by ignoring terms containing the larger hafnians (see Appendix~\ref{app:runtime}).
We test the validity of this $k$-order approximation on our experimental results further below.

\begin{figure}
    \centering
    \includegraphics[width=1\columnwidth]{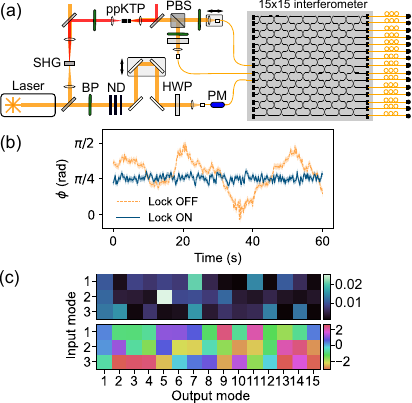}
	\caption{
	(a) Experimental setup.
	SHG: second-harmonic generation, PBS: polarizing beam splitter, BP: bandpass filter, ND: neutral-density filter, HWP: half-wave plate, PM: phase modulator.
	(b) Measurement of the phase $\phi$ between the squeezer and coherent state.
    The standard deviation of the locked curve is $\Delta \phi \sim \pi/50$.
    (c) Reconstructed transfer matrix. 
    Top (bottom) shows $|T_{ij}|^2$ ($\theta_{ij}$ in radians).
	}
	\label{fig:experimental_details}
\end{figure}

\section{Experiment}

Our experimental setup is shown in Fig.~\ref{fig:experimental_details}(a).
A mode-locked fiber laser produces pulses of 100 fs duration at a repetition rate of 10 MHz and with a center wavelength of 1550 nm.
The pulses are split into two paths.
In the top path, we frequency-double the pulses in a periodically-poled lithium niobate crystal and subsequently pump a periodically-poled potassium titanyl phosphate (ppKTP) waveguide which produces degenerate two-mode squeezed vacuum through type-II spontaneous parametric down-conversion.
In the bottom path, we prepare the coherent state.
The three beams are coupled into polarization-maintaining single-mode fibers with efficiency $\eta_c \sim 50$\% which are then coupled into a chip using grating couplers with efficiency $\eta_g \sim 70$\%.
The chip is made using silicon-on-insulator and contains a $15\times15$ network of directional couplers which comprise the interferometer.
We discuss its characterization below.
The propagation efficiency inside the 2-mm-long chip is $\eta_p \sim 70$\%.
Finally, the 15 output modes are detected using superconducting nanowire single-photon detectors.
Since these are not photon-number-resolving detectors, we can only determine ``collision-free" outcomes where $n_j \leq  1$ (see Appendix~\ref{app:collisions}).
We adjust the output light's polarization using fiber polarization controllers in every mode to maximize the detection efficiency ($\eta_d \sim 80$\%).
The total end-to-end efficiency of the experiment is $\eta_{tot} = \eta_c\eta_g^2\eta_p\eta_d \sim 10\%$.

The interference quality between the three beams depends on their modal purity and indistinguishability.
We engineer the ppKTP source to be spectrally uncorrelated and find that the modal purity of the down-converted photons is $0.85(2)$ via a second-order autocorrelation measurement.
Bandpass spectral filters are used to block the sinc-sidelobes from the down-converted spectra and to filter the classical beam. 
The temporal overlap between the three beams is controlled by two delay stages.
The single-mode nature of the on-chip directional couplers ensured spatial and polarization overlap.
Since the three beams are indistinguishable, one cannot discern whether photons detected at the output of the interferometer originated from the squeezer or the coherent state. 
Thus, the probability of detecting two or more photons depends on the relative phase between these two sources, $\phi$.
We observe $\phi$ drifting on timescales of a few seconds [orange dashed line, Fig.~\ref{fig:experimental_details}(b)] due to various instabilities in the lab.
By monitoring the twofold detection rates in real time, we construct an error signal that we then use to control the voltage applied to a phase modulator in the coherent state path and lock $\phi$ to $\pi/4$ [blue line, Fig.~\ref{fig:experimental_details}(b)].
More details on the phase locking can be found in Appendix~\ref{sec:phase_locking}.
We measure a fringe visibility of $82(2)\%$ for the two-photon interference signal obtained by combining pairs of photons from the squeezer and the coherent state on a balanced beam splitter (see Appendix~\ref{sec:source_charac}).
This visibility provides a benchmark of the overall indistinguishability and modal purity of the three beams.



The on-chip interferometer is described by a complex $15\times15$ transfer matrix $\bm{T}$.
Each element $|T_{ij}|e^{i\theta_{ij}}$ gives the probability amplitude $|T_{ij}|$ that a photon entering port $i$ exits through port $j$, while $\theta_{ij}$ is the corresponding phase.
Both quantities depend on the reflectivities and phases of the directional couplers. 
The reflectivities are chosen to follow a Haar-random distribution while the phases are randomised due to the fabrication tolerance~\cite{bell2019testing}.
Since we fix the three input modes in our experiment, we only characterize the relevant $3 \times 15$ submatrix.
The probabilities $|T_{ij}|^2$ [Fig.~\ref{fig:experimental_details}(c) top] are determined by injecting light into each input mode $i$, one at a time, and measuring the single-photon detection rates $R_{ij}$ at every output $j$.
We then normalize the rates for each input and multiply the normalized rates by the overall efficiency of the experiment, i.e. $|T_{ij}|^2 = \eta_{tot} R_{ij} /\sum_j R_{ij}$. 
The phases $\theta_{ij}$ [Fig.~\ref{fig:experimental_details}(c) bottom] are determined from the visibility of two-photon interference signals obtained by injecting two photons from the squeezer into each possible pair of inputs and recording the twofold rates.
By scanning the temporal delay between the photons, we observe Hong-Ou-Mandel-type dips whose visibilities can be related to $\theta_{ij}$~\cite{laing2012super}.


\section{Results}

\begin{figure}
    \centering
    \includegraphics[width=1\columnwidth]{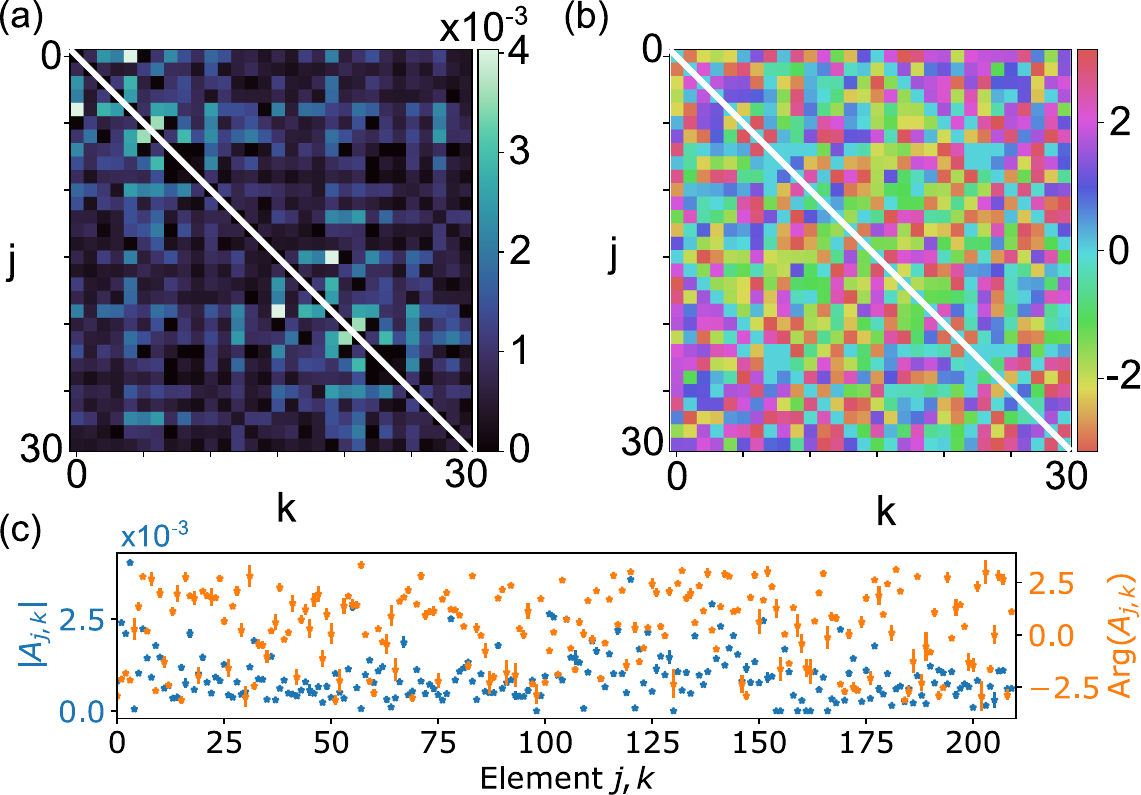}
	\caption{Measuring the 15-dimensional output Gaussian state. (a) Elements $|A_{j,k}|$ and (b) $\mathrm{arg}(A_{j,k})$ in radians of the reconstructed $\bm{A}$ matrix using the direct method described in Sec.~\ref{sec:reconstructing_state}. We cannot determine diagonal elements [white line] because our detectors are not photon-number resolving. (c) Absolute values and phases of the unique elements of $\bm{A}$ along with the associated error bars ($1\sigma$ uncertainty) obtained from the fits and Monte Carlo simulations, as explained in Appendix~\ref{sec:appendix_characterization}.}
	\label{fig:A_mat_reconstruction}
\end{figure}


We begin by demonstrating the characterization of the $\bm{A}$ matrix using two methods.
The first ``direct" method is outlined in Sec.~\ref{sec:reconstructing_state} and determines $\bm{A}$ directly from the measured photon statistics.
The second ``indirect" method calculates $\bm{A}$ by propagating a two-mode squeezed vacuum and coherent state through the measured transfer matrix $\bm{T}$ using the Python libraries Strawberryfields~\cite{killoran2019strawberry} and TheWalrus~\cite{gupt2019walrus}.
For this second method, the squeezing parameter $r$ and coherent state intensity $|\alpha|^2$ were determined by estimating the average photon numbers before losses.
Throughout the experiment, we fix the pump power at 1 mW and measured $\braket{n_{\mathrm{PDC}}} = 0.01$ photons per pulse from the squeezer, thus $r = \mathrm{arcsinh}(\sqrt{\braket{n_{\mathrm{PDC}}}/\eta_{tot}}) \sim 0.3$.
In Fig.~\ref{fig:A_mat_reconstruction}, we show the $\bm{A}$ matrix reconstructed using the direct method with $|\alpha|^2 = 1.9$.
The diagonal elements are undetermined because we do not have number-resolving detectors and thus cannot measure $p_{j,j}$ [Eq.~\eqref{eqn:twos_j_blocked}] or $p'_{j,j}$ [Eq.~\eqref{eqn:twos_j_unblocked}].
Details on the state reconstruction technique are provided in Appendix~\ref{sec:appendix_characterization}.
Since the quantum state contains mostly vacuum, metrics such as the fidelity do not provide a sensitive comparison between the $\bm{A}$ matrices obtained from both methods.
Instead, we calculate the output photon statistics of the two matrices using Eq.~\eqref{eqn:gbs_loop_haf} and compare these to the experimentally measured statistics.
We calculate $\mathrm{pr}(\bm{n}_i)$ for all 455 $N=3$ collision-free detection patterns $\bm{n}_i$ and normalize the resulting distributions, $\sum_i \mathrm{pr} (\bm{n}_i) = 1$.
The distance between the experimental and theory distributions can be computed using the total variation distance:
\begin{equation}
D = \sum_i|\mathrm{pr}_{\mathrm{exp}}(\bm{n}_i)-\mathrm{pr}_{\mathrm{th}}(\bm{n}_i)|/2,
\label{eqn:tvd}
\end{equation}
which varies between 0 and 1.
We find $D=0.033(1)$ ($D=0.0477(7)$) for the direct (indirect) method, thus showing that both methods correctly characterized $\bm{A}$.
The main advantages of the direct method are that it requires only three measurement settings and it is \textit{in situ}, i.e. it directly determines each element of $\bm{A}$ from the output statistics of the GBS machine.
In contrast, the indirect method requires injecting single photons or coherent states into every pair of input modes to determine $\bm{T}$, resulting in at least $2d-1$ measurement settings~\cite{laing2012super,rahimi2013direct}.
Then, one still has to measure the squeezing parameter of each input squeezed state and calculate how these transform under $\bm{T}$ in order to determine $\bm{A}$.

\begin{figure}
    \centering
    \includegraphics[width=1\columnwidth]{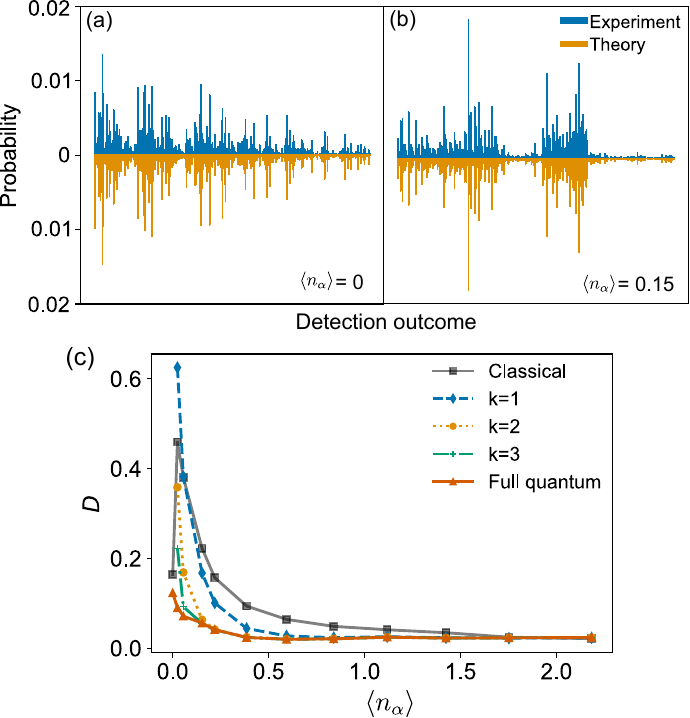}
	\caption{
	Probability distribution for all 1365 fourfold detection events when (a) $\braket{n_\alpha}=0$ and (b) $\braket{n_\alpha}=0.15$.
	Theory distributions are calculated using the full quantum model (i.e. without approximations).
    (c) Total variation distance $D$ between experiment and theory as a function of $\braket{n_\alpha}$.
    Each line corresponds to a different theory model that is further discussed in the main text.
    Error bars from Poisson counting statistics are smaller than the marker size.
    }
	\label{fig:experimental_results}
\end{figure}



Next, with $\braket{n_{\mathrm{PDC}}} = 0.01$ fixed, we increase the coherent state intensity such that its measured value $\braket{n_{\alpha}}=\eta_{tot}|\alpha|^2$ varies from 0 to 2.2.
We record the photon statistics for each value for one hour.
In Fig.~\ref{fig:experimental_results}(a) and (b), we show all 1365 measured collision-free fourfold probabilities for $\braket{n_{\alpha}}=0$ and $\braket{n_{\alpha}}=0.15$, respectively.
As before, we quantify the discrepancy between experiment and theory by calculating $D$.
We plot $D$ [red triangles, Fig.~\ref{fig:experimental_results}(c)] as a function of $\braket{n_{\alpha}}$ and find a mean of 0.04(3) with a maximum value of 0.123(4) occurring at $\braket{n_{\alpha}}=0$.
The trend of increasing $D$ for small $\braket{n_{\alpha}}$ is likely due to slight distinguishability between the down-converted modes that has a more pronounced effect when it is more probable that the photons originated from the squeezer.
We observe a similar trend for the distances obtained with the twofold and threefold distributions.
We find an average $D$ of 0.030(6) and 0.030(10) with a maximum value of 0.0421(1) and 0.0477(7) (also occurring at $\braket{n_{\alpha}}=0$) for the twofold and threefold distributions, respectively.

We also use the data collected above to study the validity of various approximate models.
We first consider the ``classical" model devised in Ref.~\cite{qi2020regimes}.
Its strategy is to determine the displaced squeezed thermal state having a classical quasiprobability distribution that best approximates the experimentally prepared state (see Appendix~\ref{app:classical_model}).
The resulting photon-number distribution can be efficiently sampled using classical algorithms~\cite{rahimi2016sufficient}.
For low $\braket{n_{\alpha}}$, the classical model has a large $D$ [black squares] and performs far worse than the quantum model.
However, for large $\braket{n_{\alpha}}$, we find that $D$ for the classical model is nearly equal to the quantum model, thus showing the classical model is a valid approximation in this regime.
This is expected because it is more likely to detect photons originating from the coherent state than the squeezer at large $\braket{n_{\alpha}}$.
The kink at $\alpha=0$ is likely an artifact of not including distinguishability in the model and thus not sampling from the optimal classical state, i.e. $D$ can be further reduced for $\alpha > 0$ (see Appendix~\ref{app:classical_model}).
Between the classical and quantum models, we can make use of semiclassical $k$-order approximations discussed in Sec.~\ref{sec:k-order}.
As we increase $\braket{n_{\alpha}}$, these approximations become better at modeling the data, as expected.

\begin{figure}
    \centering
    \includegraphics[width=1\columnwidth]{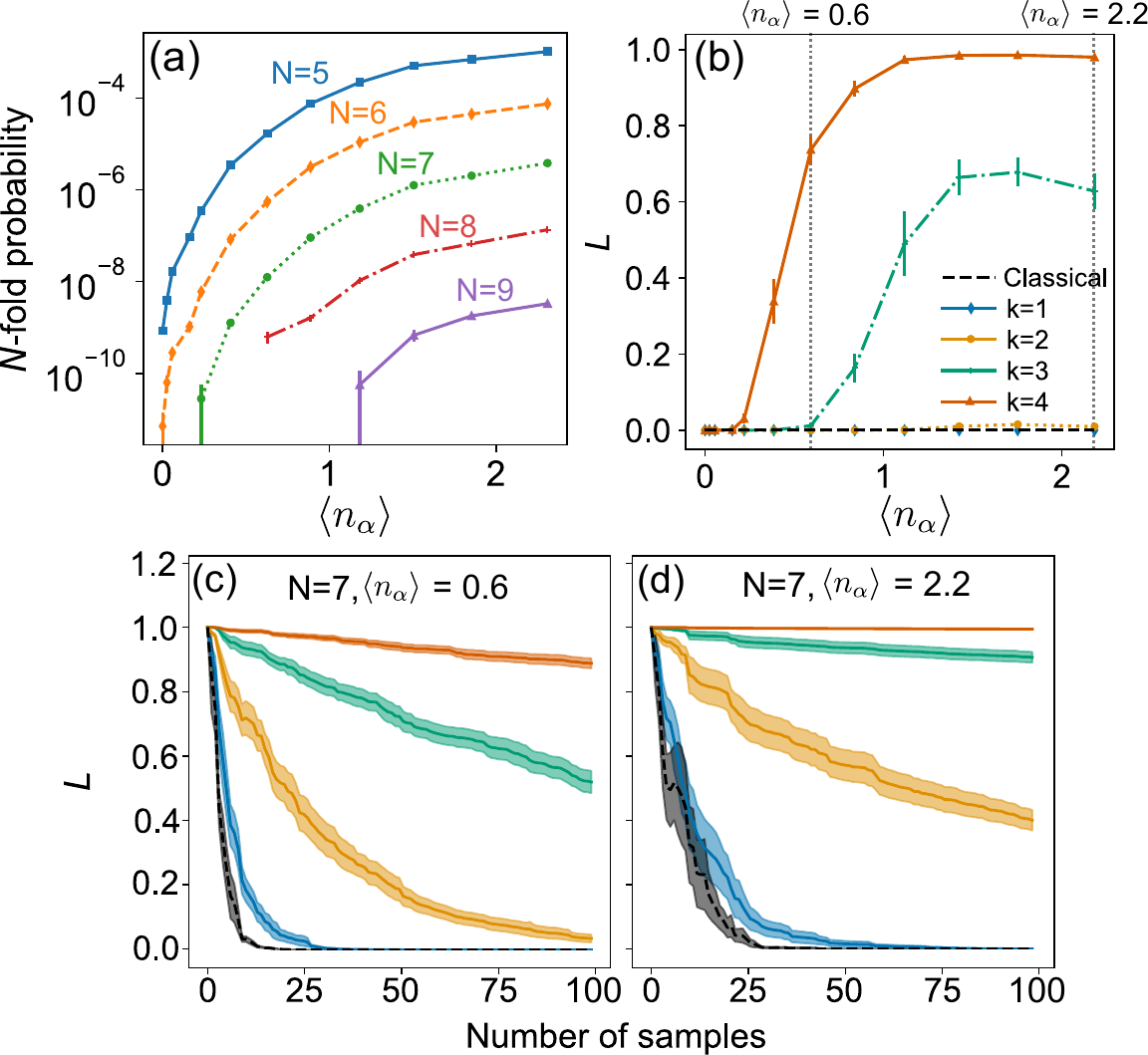}
	\caption{(a) Probability to measure a $N$-fold sample per pulse as a function of $\braket{n_\alpha}$.
	Error bars show $1\sigma$ uncertainty due to Poisson counting statistics.
	(b)-(d) Likelihood ratio $L$ [Eq.~\eqref{eqn:bayes_test}] compares the approximate model to the full quantum model.
	$L=1$ means both are equally likely.
	In (b), $L$ is calculated using $N\geq4$ samples drawn randomly from our measurement results.
	In (c) and (d), we show $L$ being updated with each new randomly drawn $N=7$ sample for $\braket{n_\alpha}=0.7$ and $2.2$, respectively.
	The lines follow the same color legend as in (b).
	Error bars in (b) and shaded regions in (c) and (d) show $1\sigma$ uncertainty obtained by repeating the calculation 10 times.
    }
	\label{fig:bayes_test}
\end{figure}

Increasing $\braket{n_{\alpha}}$ also increases the rate at which we obtain larger $N$ detection events [Fig.~\ref{fig:bayes_test}(a)].
With $\braket{n_{\alpha}}=0$, we measure fivefold events at a rate of roughly $10^{-2}$~Hz, whereas this increases to $10^{4}$~Hz with $\braket{n_{\alpha}}=2.2$.
When gauging the ability of models to predict large $N$ samples, it is more practical to use a method which does not require calculating the distance between the entire distributions as in Eq.~\eqref{eqn:tvd}.
To this end, we perform a likelihood test which compares two models $\mathcal{A}$ and $\mathcal{B}$ via the likelihood ratio 
\begin{equation}
\label{eqn:bayes_test}
    L = \prod_{i=1}^P \frac{\mathrm{pr}(\bm{n}_i|\mathcal{A})}{\mathrm{pr}(\bm{n}_i|\mathcal{B})}.
\end{equation}
Suppose $\mathcal{S}=\{\bm{n}_1,...,\bm{n}_P \}$ is a set of $P$ measured samples.
It follows from Bayes theorem that $L<1$ occurs if $\mathrm{pr}(\mathcal{B}|\mathcal{S}) >\mathrm{pr}(\mathcal{A}|\mathcal{S})$, meaning it is more likely the samples came from the probability distribution of model $\mathcal{B}$. 
We fix model $\mathcal{B}$ to be the full quantum model without approximations and calculate $L$ for various approximate models $\mathcal{A}$.
In Fig.~\ref{fig:bayes_test}(b), we plot $L$ for a sample set $\mathcal{S}$ of $P=500$ randomly chosen samples from the experimentally collected data containing at least four-photon detection events.
We find that $L$ of the $k=3,4$ models increases with $\braket{n_{\alpha}}$, and thus, as before, the validity of these models is improving as it becomes more likely that the detected photons originated from the coherent state.
We also show the trend of $L$ with each new $N=7$ sample [Fig.~\ref{fig:bayes_test}(c),(d)].
Despite the larger number of photons detected, the $k=4$ approximation still appears to be valid.
However, unlike in Fig.~\ref{fig:experimental_results}(c), the $k\leq 2$ approximations appear to be inadequate to model the $N\geq4$ data regardless of $\braket{n_{\alpha}}$.
This suggests that these approximate models cannot accurately predict higher-order correlations even at high $\braket{n_{\alpha}}$.

\section{Conclusions}
We experimentally implemented a GBS machine that samples from a displaced nonclassical Gaussian state.
We introduced and tested the validity of approximate semiclassical models that exploit the classical nature of the displacement to speed up calculations when this quantity is large relative to the squeezer strength. 
Moreover, we showed that the displacement field enables the reconstruction of the Gaussian state at the output of a GBS machine using only three measurement settings.
The techniques introduced here will be useful for characterizing and validating large-scale GBS experiments.
In particular, the ability to efficiently reconstruct the output Gaussian state can be used to verify that the degree of squeezing and displacement as well as the interferometer transformation has been correctly set for a desired calculation. 
Moreover, as with approximate models that exploit experimental imperfections in sources and detectors to speed up GBS calculations~\cite{renema2018efficient,garcia2019simulating,qi2020regimes,bulmer2022boundary}, the $k$-order models we introduced exploit the classical contribution of the displacement field.
These various models can be used together to better gauge the computational difficulty of sampling the output light distribution of a GBS machine.

We briefly comment on the prospect of using GBS with displacement for simulating molecular vibronic spectra.
The required displacement energy varies widely depending on the molecule, and can even be significantly larger than the squeezed vacuum energy.
For example, simulating the vibronic spectra of formic acid~\cite{huh2015boson} (sulfur dioxide~\cite{huh2017vibronic}) uses roughly 0.07 (0.014) photons from squeezers and a displacement of about 1.5 (1.6) photons, whereas certain transitions in tropoline requires only squeezing and no displacement~\cite{clements2018approximating}.
Although these numbers were achievable in our setup, the covariance matrix and displacement of the output Gaussian state was fixed by the static interferometer.
We provide a recipe to implement arbitrary transformations using a reconfigurable multiport interferometer in Appendix~\ref{app:arb_disp}, which could simulate many molecules in a single GBS machine.
Moreover, losses will reduce the fidelity of the simulated spectra, but this can be partially mitigated by optimizing the displacement and squeezing~\cite{clements2018approximating, banchi2020molecular}.
Finally, we also note that GBS can inspire more efficient classical algorithms for calculating vibronic spectra~\cite{quesada2019franck,oh2022quantum}.
In particular, our k-order approximations could be useful to simulate systems having a large displacement energy relative to the squeezing.

\begin{acknowledgements}
We thank Renyou Ge and Xinlun Cai for fabricating the silicon chip.
We also thank Jacob Bulmer and Gabriele Bressanini for their comments on the manuscript.
This work is supported by the Engineering and Physical Sciences Research Council (P510257 and T001062), H2020 Marie Sklodowska-Curie Actions (846073), Samsung GRC, and the KIST Open Research Program.
\end{acknowledgements}

\appendix


\section{Source characterization}
\label{sec:source_charac}
\begin{figure}
    \centering
    \includegraphics[width=1\columnwidth]{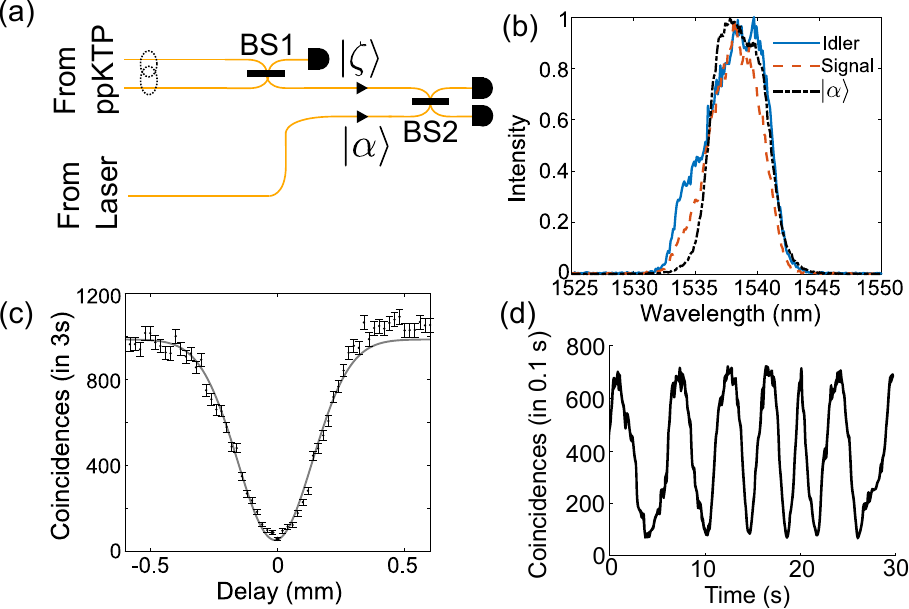}
	\caption{
	(a) Setup for measuring interference between the squeezer and coherent state.
	(b) Spectrum of three interfering modes.
    (c) Hong-Ou-Mandel dip between down-converted modes. 
    $\mathcal{V} = 94(4)\%$.
    (d) Interference signal between $\ket{\zeta}$ and $\ket{\alpha}$.
	$\mathcal{V} = 82(2)\%$. The phase $\phi$ between the two drifts randomly for this measurement.
	}
	\label{fig:source_characterization}
\end{figure}

We use two-photon interference between the squeezer and coherent state to obtain a benchmark of the overall quality of the indistinguishability and modal purity of the three modes [Fig.~\ref{fig:source_characterization}(a),(b)].
We first combine the two down-converted modes on a balanced beam splitter (BS1).
Because of Hong-Ou-Mandel interference, the down-converted photons bunch and thus we observe a dip in coincidences at the BS1 output of visibility  $\mathcal{V}=94(4)\%$ [Fig.~\ref{fig:source_characterization}(c)].
Consequently, light in the bottom output port of BS1 is approximately in a single-mode squeezed vacuum state $\ket{\zeta}$.
On BS2, we combine $\ket{\zeta}$ with a coherent state $\ket{\alpha}$ whose amplitude is set such that the two-photon probability is roughly equal to that of the squeezed vacuum state, i.e. $\left|\braket{2|\zeta}\right|^2 \approx \left|\braket{2|\alpha}\right|^2$.
By measuring coincidence events at the output of BS2, we observe a two-photon interference signal with $\mathcal{V}=82(2)\%$ [Fig.~\ref{fig:source_characterization}(d)].
For our experimental parameters ($r\sim 0.3$, $|\alpha|^2 \sim 0.3$, $\eta \sim 0.4$), we numerically calculated that the upper limit on this visibility is $94\%$.
The ratio of our measured visibility to the ideal one is consistent with the modal purity 0.85(2) of the down-converted light determined via a second-order autocorrelation measurement.

\section{Phase locking}
\label{sec:phase_locking}

\begin{figure}
    \centering
    \includegraphics[width=1\columnwidth]{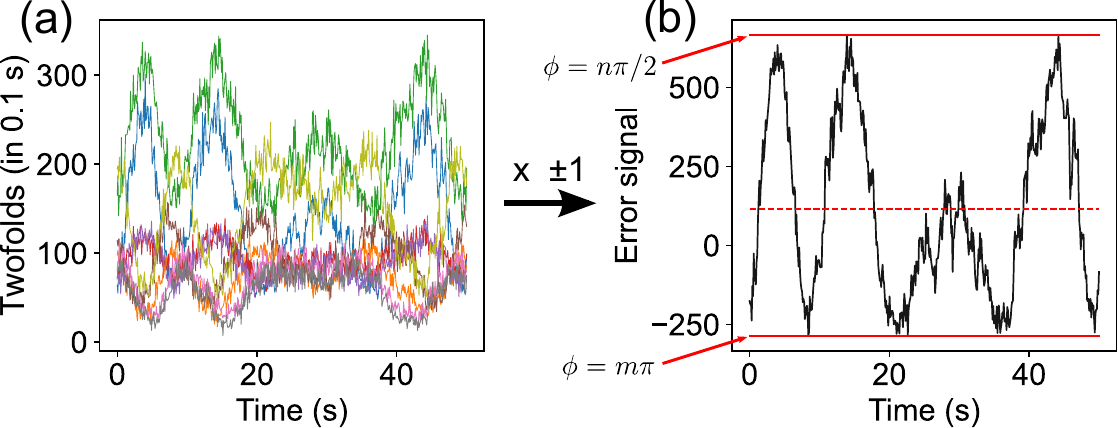}
	\caption{
	(a) A subset of the twofold detection rates $p_{j,k}(\phi)$ used in the error signal to lock the phase $\phi$. (b) The error signal obtained by adding the twofold detection rates with the appropriate sign. $n$ is a non-zero integer while $m$ is any integer. 
	}
	\label{fig:error_signal}
\end{figure}

We use twofold photon statistics to lock the phase $\phi$ between the squeezer and the coherent state.
The probability to measure a photon in output mode $j$ and $k$ is given by Eq.~\eqref{eqn:twos_j_unblocked}.
The last term in this equation is an interference term that depends on $\phi$.
The visibility of this interference depends on the relative likelihood that the coherent state and squeezer each produced two photons and that these photons exit the circuit in output modes $j$ and $k$.
We construct an error signal by heuristically choosing (i.e. those with a high interference visibility) pairs of $j,k$ and add their respective twofold rates $p_{j,k}(\phi)$. 
A subset of these rates is shown in Fig.~\ref{fig:error_signal}(a).
The rates are either correlated or anticorrelated with respect to one another depending on whether Hong-Ou-Mandel bunching or antibunching occurs, which depends on the internal phases of the interferometer.
The error signal shown in Fig.~\ref{fig:error_signal}(b) is obtained by summing these rates with the anticorrelated ones multiplied by $-1$.
We use this error signal in a PID loop in order to control the voltage of the phase modulator and lock $\phi$ to $\pi/4$.
The voltage is updated every 0.1s.

\section{Gaussian state characterization}
\label{sec:appendix_characterization}
In Sec.~\ref{sec:reconstructing_state}, we showed that by controlling the phase of a coherent state injected in one input mode of the interferometer and measuring single-photon and two-photon detection probabilities, we can nearly fully characterize the output Gaussian state. 
The only missing quantity is the sign of the imaginary part of $C_{j,k}$.
Here we show how to determine this sign by injecting the coherent state into a second input mode.

\begin{figure}[t]
    \centering
    \includegraphics[width=1\columnwidth]{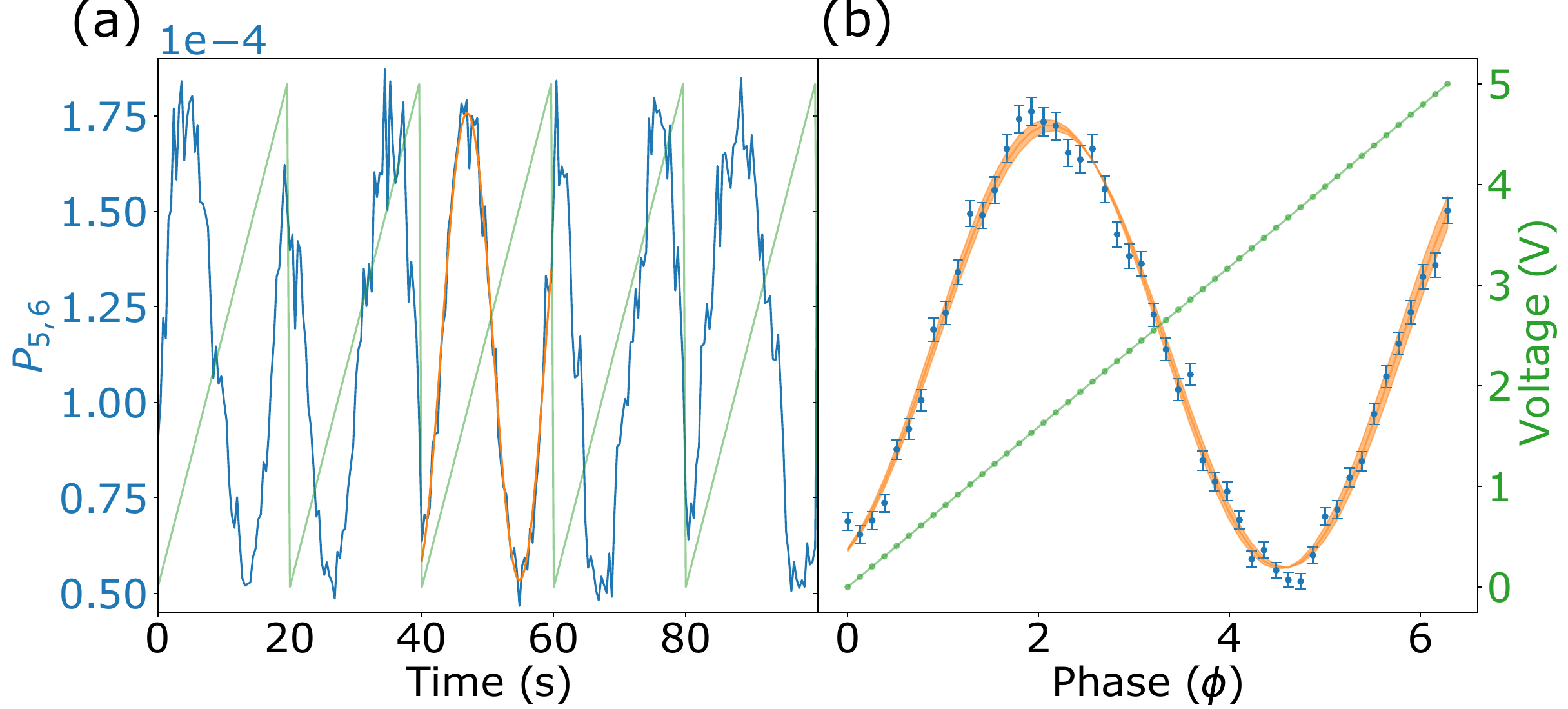}
	\caption{
	(a) Probability of a twofold detection between modes 5 and 6 while sweeping the phase $\phi$. Orange curve shows the fit of Eq.~\eqref{eqn:twos_j_unblocked} in a $2\pi$ region where the fitting error is minimized. Green line indicates the phase modulator voltage value. The twofold rate for this example was approximately 750 per second. (b) Twofold probabilities as a function of $\phi$. Error bars and shaded region show $1\sigma$ uncertainty, obtained from Poissonian counting statistics and variances in the fitting, respectively.
	}
	\label{fig:probability_osc}
\end{figure}

With the coherent state injected in a first input mode, Eqs.~\eqref{eqn:singles_j_unblocked} and ~\eqref{eqn:twos_j_unblocked} provide the single-photon and two-photon probabilities.
In these equations, we assumed the elements of $\bm{\gamma}$ to be real valued, thus fixing a phase reference in the output of the interferometer.
Injecting the coherent state into a different input mode, we obtain the analogous equations
\begin{subequations}
\begin{align}
    p''_j &= C_{j,j} + |\mu_j|^2, \label{eqn:singles_j_unblocked_prime}\\
\begin{split} 
    p''_{j,k} &= p''_jp''_k + |B_{j,k}|^2 + |C_{j,k}|^2 \\ 
    &\quad + 2 \left(\mathrm{Re}[\mu_j\mu_k^* C_{j,k}] + \mathrm{Re}[\mu_j\mu_k B^*_{j,k}e^{i2\phi}] \right). 
\end{split}
\label{eqn:twos_j_unblocked_prime}
\end{align}
\end{subequations}
where the elements of $\bm{\mu}$ are now complex valued.
We can determine the absolute value $|\mu_j|$ via Eq.~\eqref{eqn:singles_j_unblocked_prime} by using the already known $C_{j,j}$.
As before, the last term in Eq.~\eqref{eqn:twos_j_unblocked_prime} is an interference term leading to a fringe that can be observed by scanning the phase $\phi$ of the coherent state [Fig.~\ref{fig:probability_osc}].
The phase offset of this fringe can be used to determine $\mathrm{arg}(\mu_k)$ since $\mathrm{arg}(B_{j,k})$ is already known and we are free to choose one of the output phases of $\bm{\mu}$, e.g. $\mathrm{arg}(\mu_1)=0$.
This assumes that the coherent state injected in the second input mode has the same phase $\phi$ as when injected in the first input.
If instead there is an unknown offset between the two phases, one can set $\mathrm{arg}(\mu_1)=\tilde{\phi}$ and solve for this single unknown parameter by minimizing the distance for the threefold photon statistics [Eq.~\eqref{eqn:tvd}].
The imaginary part of $C_{j,k}$ is then determined by
\begin{equation}
\begin{split}
\mathrm{Im}[C_{j,k}] &= \epsilon_{j,k} \mathrm{Re}(C_{j,k})[\mathrm{Re}(\mu_j) \mathrm{Re}(\mu_k) + \mathrm{Im}(\mu_j) \mathrm{Im}(\mu_k)] \\
&\quad - \epsilon_{j,k} \mathrm{Re}[\mu_j\mu_k^* C_{j,k}], \\
\end{split}
\end{equation}
where $\epsilon_{j,k} = \left(\mathrm{Im}(\mu_j)\mathrm{Re}(\mu_k)-\mathrm{Re}(\mu_j)\mathrm{Im}(\mu_k)\right)^{-1}$ and $\mathrm{Re}[\mu_j\mu_k^* C_{j,k}]$ is obtained from Eq.~\eqref{eqn:twos_j_unblocked_prime}.

Fig.~\ref{fig:probability_osc} shows an example of this procedure for a particular pair of modes, $(j,k)=(5,6)$.
We first collect data while sweeping the phase modulator voltage [Fig.~\ref{fig:probability_osc}(a)].
We then fit Eq.~\eqref{eqn:twos_j_unblocked} in $2\pi$ regions of the phase scan.
The final fit is obtained by averaging over the five regions with the smallest fitting errors to minimize the effect of phase fluctuations and reduce Poissonian counting statistic errors [Fig.~\ref{fig:probability_osc}(b)].
Fitting errors are propagated through Eqs.~\eqref{eqn:singles_j_unblocked} and~\eqref{eqn:twos_j_unblocked} to determine the uncertainty on the recovered matrix elements $B_{j,k}$ and $C_{j,k}$.

In theory, the optimal displacement value maximizes the amplitude of the oscillating term in Eq.~\eqref{eqn:twos_j_unblocked}.
In practice, since we are using non-number-resolving detectors, we employ a weaker displacement of $\braket{n_\alpha}=0.19$ to reduce the effect of collisions (see Appendix~\ref{app:collisions}).
Four output mode pairs produced near-zero twofold detection rates (i.e. about 1 per second) due to very low transmission probabilities through the interferometer.
Fitting these rates with Eq.~\eqref{eqn:twos_j_unblocked} leads to a near zero $|B_{j,k}|$ and $|C_{j,k}|$, as expected, but also leaves the phase of these matrix elements undetermined.
Other effects such as instabilities in the phase $\phi$ or counting statistics errors can also hinder the fitting when the twofold rates are low.
A detailed study of the robustness and limitations of our reconstruction method will be presented in a future work. 

To resolve these issues here, we determine the phases that minimize the distance [Eq.~\eqref{eqn:tvd}] for the threefold distribution using a numerical optimization algorithm.
The error on the resulting phases is determined via a Monte Carlo approach: we run the optimization ten times using a different set of initial values for the phases on each run.
The initial values are obtained by sampling from a Gaussian distribution of mean $\mathrm{arg}(A_{j,k})$ and standard deviation given by the corresponding uncertainty. 
For the phases of the four elements that could not be retrieved with the direct inversion, we sampled from a uniform distribution between $[-\pi, \pi)$.

\section{Collisions}
\label{app:collisions}
Our experiment employs ``click" detectors that cannot resolve photon numbers.
As such, events in which an output mode contained more than one photon, $n_j > 1$, are convolved in the measured probabilities of the collision-free events.
This leads to an error in the estimate of the collision-free probabilities.


To estimate the relative size of errors caused by collisions, we calculate the probability of a collision-free event $\bm{n}$ using the loop Torontonian~\cite{bulmer2022threshold}:
\begin{equation}
\mathrm{pr}(\bm{n}) = p_{\mathrm{vac}}\times \mathrm{ltor}( \tilde{\bm{A}}_{\bm{n}} ).
\label{eqn:ltor}
\end{equation}
Unlike the loop hafnian [Eq.~\eqref{eqn:gbs_loop_haf}], the loop Torontonian determines the photon statistics measured by click detectors, i.e. it convolves the probabilities of collision events with $n_j > 1$.
An implementation of Eq.~\eqref{eqn:ltor} and Eq.~\eqref{eqn:gbs_loop_haf} can be found in the Python package TheWalrus~\cite{gupt2019walrus}.
We compute the distance $D$ between the distributions obtained using the two equations for all collision-free $N=4$ detection outcomes.
We find that $D$ increases with $\braket{n_\alpha}$ with a maximum of $D=0.024$ occurring at $\braket{n_{\alpha}}=2.2$.
Thus, the error caused by collisions is relatively small.  

\section{Classical model}
\label{app:classical_model}
The classical model, devised in Ref.~\cite{qi2020regimes}, determines the displaced squeezed thermal state having a classical quasiprobability distribution with the highest fidelity (i.e. state overlap) to the experimentally prepared GBS state.
One can then calculate its photon statistics using classical algorithms such as those presented in Ref.~\cite{rahimi2016sufficient}.
To determine this classical state, we follow Algorithm 1 given in the Supplementary Material of Ref.~\cite{qi2020regimes}, which we reproduce here.

We begin by finding the classical squeezed thermal state that approximates the down-converted light given the total end-to-end efficiency of our experiment, $\eta$.
After losses, a squeezed vacuum state with squeezing parameter $r$ is transformed to a squeezed thermal state whose covariance matrix is given by $\bm{V}=\mathrm{diag}(a_+, a_-)$ where $a_\pm = \eta e^{\pm2r}+(1-\eta)$.
Using our experimental parameters ($\eta\sim0.1$, $r \sim 0.28$), this covariance matrix is nonclassical since $\bm{V}-\bm{I}_2$ is not positive semidefinite.
The closest classical state is a squeezed thermal state with squeezing parameter $s$ and thermal occupation number $n$~\cite{marian2002quantifying}:
\begin{subequations}
\begin{align}
    n &= -\frac{1}{2} + \frac{1}{2}\sqrt{1+2\sinh(2s_c)\sqrt{a_+/a_-}}\\
    s &= \frac{1}{2} \ln\left(2n+1\right)
    \end{align}
\end{subequations}
with $s_c = \ln(\sqrt{a_+a_-})$. We propagate two such squeezed thermal states and a coherent state of intensity $|\alpha|^2$ through the interferometer using Strawberryfields~\cite{killoran2019strawberry}.
Since our down-conversion source produces two-mode squeezed vacuum, we interfere both squeezed thermal states on a fictitious balanced beam splitter before the interferometer. 

\begin{figure}
    \centering
    \includegraphics[width=0.6\columnwidth]{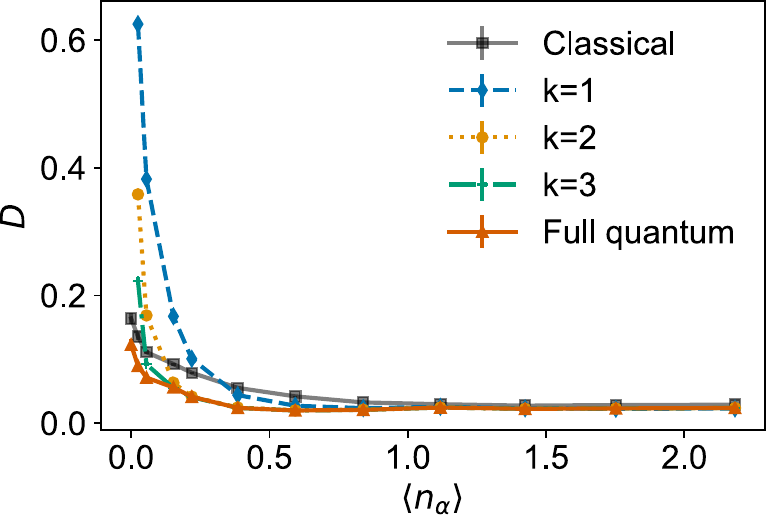}
	\caption{
	Distances $D$ of fourfold distributions obtained using different theory models. In contrast to Fig.~\ref{fig:experimental_results}(c), here the parameters of the classical model are optimized to minimize $D$.
	}
	\label{fig:distances_optimized_classical}
\end{figure}

For the sake of comparing different models in Fig.~\ref{fig:experimental_results}(c), we do not optimize the measured parameters $|\alpha|^2, \eta, r$ to minimize the distance of the classical model, i.e. we use the same parameters for all models.
To gauge the best possible performance of the classical model, we perform this optimization in Fig.~\ref{fig:distances_optimized_classical}. 
The distance is further reduced compared to Fig.~\ref{fig:experimental_results}(c) likely because distinguishability is not included in the model.

\section{Arbitrary transformation}
\label{app:arb_disp}

\begin{figure}
    \centering
    \includegraphics[width=0.7\columnwidth]{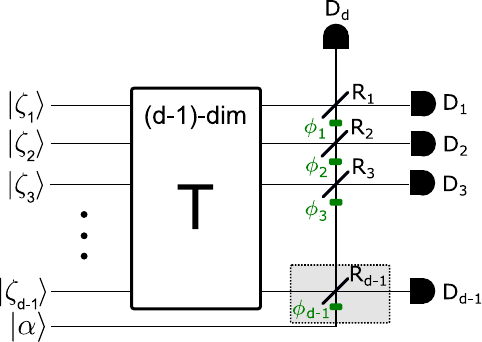}
	\caption{Recipe for arbitrary displacements using a single input coherent state.}
	\label{fig:arbitrary_disp}
\end{figure}

In our experiment, the output state's displacement and squeezing is fixed by the static interferometer.
If one instead uses a reconfigurable multiport interferometer capable of implementing any $d$-dimensional unitary transformation, then a more general $(d-1)$-dimensional Gaussian state can be prepared using the recipe shown in Fig.~\ref{fig:arbitrary_disp}.
Such a multiport interferometer contains at least $d(d-1)/2$ tunable beam splitters~\cite{clements2016optimal}.
The displacement operation $\hat{D}(\delta_j)$ is achieved by combining each output mode $j$ with the coherent state on a beam splitter of low reflectivity $R_{j}\ll 1$ and phase shift $\phi_j$~\cite{paris1996displacement}.
This leaves $(d-1)(d-2)/2$ beam splitters for the squeezers, which can be used for an arbitrary $(d-1)$-dimensional unitary.
We also note that setting $\bm{T}$ to the identity and measuring coincidences between detector $D_d$ and each $D_j$ can be used to lock the phases of every squeezed vacuum to the coherent state using a procedure similar to that presented in Appendix~\ref{sec:phase_locking}.

\section{k-order run time}
\label{app:runtime}

\begin{figure}[ht]
    \centering
    \includegraphics[width=0.75\columnwidth]{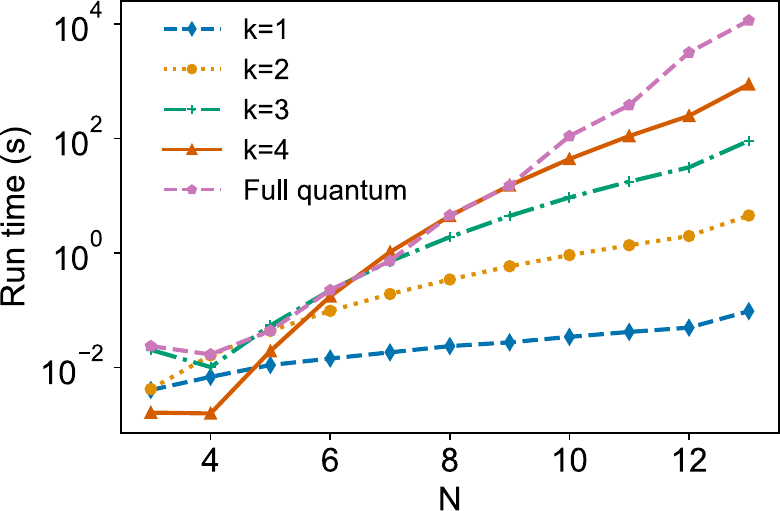}
	\caption{Run time of the exact loop hafnian [Eq.~\eqref{eqn:gbs_haf}], i.e. ``full quantum'' model, and the approximate k-order models, as a function of the number of detected photons $N$.}
	\label{fig:run_time}
\end{figure}

In Fig.~\ref{fig:run_time}, we plot the run time of a typical calculation of the loop hafnian [Eq.~\eqref{eqn:gbs_haf}] and the k-order approximation which truncates this quantity at a certain $k$ (see Sec.~\ref{sec:k-order}).
The calculations are performed on a desktop machine with a 16-core 2.9 GHz CPU and 16 GB of memory.
The loop hafnian is calculated using the TheWalrus~\cite{gupt2019walrus} whereas the k-order approximation uses our own code (available upon request).
While the former code has been well optimized~\cite{bulmer2022boundary}, our k-order approximation implementation is likely not optimal and we anticipate that its run time can be further improved.

\bibliographystyle{apsrev4-2}
\bibliography{refs}


\end{document}